\documentclass[twocolumn]{aastex63}

\usepackage{xspace}
\usepackage{enumitem}
\usepackage[utf8]{inputenc}
\usepackage{newunicodechar,graphicx}
\usepackage{amsmath}
\usepackage{lineno}
\usepackage{tabularx}
\usepackage{booktabs}
\usepackage{upgreek}
\usepackage{longtable}
\usepackage{footnote}
\usepackage[T1]{fontenc}
\usepackage{CJK}

\DeclareRobustCommand{\okina}{%
  \raisebox{\dimexpr\fontcharht\font`A-\height}{%
    \scalebox{0.8}{`}%
  }%
}
\newunicodechar{ʻ}{\okina}

\newcommand{\tess}{\textit{TESS}\xspace}
\newcommand{\ktwo}{\textit{K2}\xspace}
\newcommand{\kepler}{\textit{Kepler}\xspace}

\newcommand{\lightkurve}{\textsf{lightkurve}\xspace}
\newcommand{\exoplanet}{\textsf{exoplanet}\xspace}

\newcommand{\giants}{\textsf{giants}\xspace}

\newcommand{\vsini}{\mbox{$v\sin i$}\xspace}

\newcommand{\starname}{TOI-7041\xspace}
\newcommand{\planetname}{TOI-7041 b\xspace}
\newcommand{\mstar}{$1.07\pm0.05\mathrm{(stat)}\pm0.02\mathrm{(sys)}$\xspace}
\newcommand{\rstar}{$4.10\pm0.06\mathrm{(stat)}\pm0.05\mathrm{(sys)}$\xspace}
\newcommand{\starage}{$10.3\pm1.9\mathrm{(stat)}\pm0.1\mathrm{(sys)}$\xspace}
\newcommand{\teff}{$4700\pm100$\xspace}
\newcommand{\logg}{$3.244\pm0.007\mathrm{(stat)}\pm0.001\mathrm{(sys)}$\xspace}
\newcommand{\feh}{$0.25\pm0.10$\xspace}
\newcommand{\porb}{$9.691\pm0.006$\xspace}
\newcommand{\mplanet}{$0.36\pm0.16$\xspace}
\newcommand{\rplanet}{$1.02\pm0.03$\xspace}
\newcommand{\rprs}{$0.0256\pm0.0006$\xspace}
\newcommand{\ecc}{$0.04\pm0.04$\xspace}
\newcommand{\impactparam}{$0.2\pm0.1$\xspace}

\newcommand{\plomega}{$131\pm52$\xspace}
\newcommand{\rvk}{$36.2\pm5.0$\xspace}
\newcommand{\snumax}{$218.50 \pm 2.23$\xspace}
\newcommand{\sdnu}{$16.5282 \pm 0.0186$\xspace}
\newcommand{\sdzerotwo}{$2.0906 \pm 0.0491$\xspace}

\shorttitle{\tess GTG. VII. Hot Saturn Orbiting an Oscillating Red Giant}
\shortauthors{Saunders et al.}

\graphicspath{{./}{figures/}}

\begin{document}

\title{\tess Giants Transiting Giants. VII. A Hot Saturn Orbiting an Oscillating Red Giant Star\footnote{This paper includes data gathered with the 6.5 meter Magellan Telescopes located at Las Campanas Observatory, Chile.}}

\begin{CJK*}{UTF8}{gbsn}
\author[0000-0003-2657-3889]{Nicholas Saunders}
\altaffiliation{NSF Graduate Research Fellow}
\affiliation{Institute for Astronomy, University of Hawaiʻi at M\=anoa, 2680 Woodlawn Drive, Honolulu, HI 96822, USA}

\author[0000-0003-4976-9980]{Samuel K.\ Grunblatt}
\affiliation{William H.\ Miller III Department of Physics \& Astronomy, Johns Hopkins University, 3400 N Charles St, Baltimore, MD 21218, USA}
\affiliation{Department of Physics and Astronomy, The University of Alabama, 514 University Blvd., Tuscaloosa, AL 35487, USA}

\author[0000-0001-8832-4488]{Daniel Huber}
\affiliation{Institute for Astronomy, University of Hawaiʻi at M\=anoa, 2680 Woodlawn Drive, Honolulu, HI 96822, USA}
\affiliation{Sydney Institute for Astronomy (SIfA), School of Physics, University of Sydney, NSW 2006, Australia}

\author[0000-0001-7664-648X]{J. M. Joel Ong (王加冕)}
\altaffiliation{NASA Hubble Fellow}
\affiliation{Institute for Astronomy, University of Hawaiʻi at M\=anoa, 2680 Woodlawn Drive, Honolulu, HI 96822, USA}

\author[0000-0001-5761-6779]{Kevin C.\ Schlaufman}
\affiliation{William H.\ Miller III Department of Physics \& Astronomy, Johns Hopkins University, 3400 N Charles St, Baltimore, MD 21218, USA}

\author[0000-0003-3244-5357]{Daniel Hey}
\affiliation{Institute for Astronomy, University of Hawaiʻi at M\=anoa, 2680 Woodlawn Drive, Honolulu, HI 96822, USA}

\author[0000-0003-3020-4437]{Yaguang Li (李亚光)}
\affiliation{Institute for Astronomy, University of Hawaiʻi at M\=anoa, 2680 Woodlawn Drive, Honolulu, HI 96822, USA}

\author[0000-0003-1305-3761]{R.P. Butler}
\affiliation{Earth and Planets Laboratory, Carnegie Institution for Science, 5241 Broad Branch Rd NW, Washington, DC 20015}

\author[0000-0002-5226-787X]{Jeffrey D. Crane}
\affiliation{The Observatories of the Carnegie Institution for Science, 813 Santa Barbara Street, Pasadena, CA, 91101}

\author[0000-0002-8681-6136]{Steve Shectman}
\affiliation{The Observatories of the Carnegie Institution for Science, 813 Santa Barbara Street, Pasadena, CA, 91101}


\author[0009-0008-2801-5040]{Johanna K. Teske}
\affiliation{Earth and Planets Laboratory, Carnegie Institution for Science, 5241 Broad Branch Rd NW, Washington, DC 20015}
\affiliation{The Observatories of the Carnegie Institution for Science, 813 Santa Barbara Street, Pasadena, CA, 91101}


\author[0000-0002-8964-8377]{Samuel N. Quinn}
\affiliation{Center for Astrophysics \textbar \ Harvard \& Smithsonian, 60 Garden Street, Cambridge, MA 02138, USA}

\author[0000-0001-7961-3907]{Samuel W.\ Yee}
\altaffiliation{51 Pegasi b Fellow}
\affiliation{Center for Astrophysics \textbar \ Harvard \& Smithsonian, 60 Garden Street, Cambridge, MA 02138, USA}



\author[0000-0002-9158-7315]{Rafael Brahm}
\affiliation{Facultad de Ingenier\'ia y Ciencias, Universidad Adolfo Ib\'añez, Av. Diagonal las Torres 2640, Peñalol\'en, Santiago, Chile}
\affiliation{Millennium Institute for Astrophysics, Santiago, Chile}
\affiliation{Data Observatory Foundation, Santiago, Chile}

\author[0000-0002-0236-775X]{Trifon Trifonov} 
\affiliation{Landessternwarte, Zentrum f\"ur Astronomie der Universt\"at Heidelberg, K\"onigstuhl 12, 69117 Heidelberg, Germany}
\affiliation{Department of Astronomy, Sofia University St Kliment Ohridski, 5 James Bourchier Blvd, BG-1164 Sofia, Bulgaria}

\author[0000-0002-5389-3944]{Andr\'es Jord\'an} 
\affiliation{Facultad de Ingenier\'ia y Ciencias, Universidad Adolfo Ib\'añez, Av. Diagonal las Torres 2640, Peñalol\'en, Santiago, Chile}
\affiliation{Millennium Institute for Astrophysics, Santiago, Chile}
\affiliation{Data Observatory Foundation, Santiago, Chile}


\author{Thomas Henning} 
\affiliation{Max-Planck-Institut f\"ur Astronomie, Königstuhl 17, 69117 Heidelberg, Germany}

\author[0000-0001-6050-7645]{David K.\ Sing}
\affiliation{William H.\ Miller III Department of Physics \& Astronomy, Johns Hopkins University, 3400 N Charles St, Baltimore, MD 21218, USA}

\author[0000-0001-7891-8143]{Meredith MacGregor}
\affiliation{William H.\ Miller III Department of Physics \& Astronomy, Johns Hopkins University, 3400 N Charles St, Baltimore, MD 21218, USA}

\author[0000-0002-3221-3874]{Emma Page}
\affiliation{Department of Physics, Lehigh University, 16 Memorial Drive East, Bethlehem, PA 18015, USA}

\author[0000-0003-2196-6675]{David Rapetti}
\affiliation{NASA Ames Research Center, Moffett Field, CA 94035, USA}
\affiliation{Research Institute for Advanced Computer Science, Universities Space Research Association, Washington, DC 20024, USA}

\author{Ben Falk} 
\affiliation{Space Telescope Science Institute, 3700 San Martin Drive, Baltimore, MD, 21218, USA}


\author[0000-0001-8172-0453]{Alan~M.~Levine} 
\affiliation{Department of Physics and Kavli Institute for Astrophysics and Space Research, Massachusetts Institute of Technology, Cambridge, MA 02139, USA}

\author[0000-0003-0918-7484]{Chelsea X.\ Huang}
\affiliation{University of Southern Queensland, Centre for Astrophysics, West Street, Toowoomba, QLD 4350, Australia}

\author[0000-0003-2527-1598]{Michael B.\ Lund}
\affiliation{NASA Exoplanet Science Institute, IPAC, California Institute of Technology, Pasadena, CA 91125, USA}

\author[0000-0003-2058-6662]{George~R.~Ricker}
\affiliation{Department of Physics and Kavli Institute for Astrophysics and Space Research, Massachusetts Institute of Technology, Cambridge, MA 02139, USA}



\author[0000-0002-6892-6948]{S.~Seager}
\affiliation{Department of Physics and Kavli Institute for Astrophysics and Space Research, Massachusetts Institute of Technology, Cambridge, MA 02139, USA}
\affiliation{Department of Earth, Atmospheric and Planetary Sciences, Massachusetts Institute of Technology, Cambridge, MA 02139, USA}
\affiliation{Department of Aeronautics and Astronautics, MIT, 77 Massachusetts Avenue, Cambridge, MA 02139, USA}

\author[0000-0002-4265-047X]{Joshua~N.~Winn}
\affiliation{Department of Astrophysical Sciences, Princeton University, 4 Ivy Lane, Princeton, NJ 08544, USA}

\author[0000-0002-4715-9460]{Jon~M.~Jenkins}
\affiliation{NASA Ames Research Center, Moffett Field, CA 94035, USA}

\begin{abstract}
    We present the discovery of \planetname (TIC 201175570 b), a hot Saturn transiting a red giant star with measurable stellar oscillations. We observe solar-like oscillations in \starname with a frequency of maximum power of $\nu_{\rm max}=$ \snumax $\mu$Hz and a large frequency separation of $\Delta\nu=$ \sdnu $\mu$Hz. Our asteroseismic analysis indicates that \starname has a radius of \rstar $R_\odot$, making it one of the largest stars around which a transiting planet has been discovered with the \textit{Transiting Exoplanet Survey Satellite} (\tess), and the mission's first oscillating red giant with a transiting planet. \planetname has an orbital period of \porb days and a low eccentricity of $e=$ \ecc. We measure a planet radius of \rplanet $R_J$ with photometry from \tess, and a planet mass of \mplanet $M_J$ ($114\pm51$ $M_\oplus$) with ground-based radial velocity measurements. \planetname appears less inflated than similar systems receiving equivalent incident flux, and its circular orbit indicates that it is not undergoing tidal heating due to circularization. The asteroseismic analysis of the host star provides some of the tightest constraints on stellar properties for a \tess planet host and enables precise characterization of the hot Saturn. This system joins a small number of \tess-discovered exoplanets orbiting stars that exhibit clear stellar oscillations and indicates that extended \tess observations of evolved stars will similarly provide a path to improved exoplanet characterization.
\end{abstract}

\section{Introduction} \label{sec:intro}

Asteroseismology, the study of stellar oscillations, provides uniquely precise constraints on stellar properties. The power of asteroseismology for characterizing transiting planet hosts was made evident by the space-based observatories \textit{CoRoT} \citep{ballot11b,lebreton2014a} and \textit{Kepler} \citep{huber13b,lundkvist16,vaneylen18}. The \ktwo mission expanded the sample of oscillating planet hosts and provided key constraints on the properties of evolved stars and their planets \citep{grunblatt19}. By revealing precise stellar masses, radii, and ages, these missions enabled the best characterization of transiting exoplanets. 

The \textit{Transiting Exoplanet Survey Satellite} (\tess, \citealt{Ricker2015JATIS...1a4003R}) has dramatically expanded the sample of potential targets that can benefit from the synergy between asteroseismology and exoplanet science. According to pre-launch predictions, hundreds of solar-like oscillator planet hosts (primarily low-luminosity red giant branch stars) would be detected in \tess photometry \citep{campante16}. The discovery of TOI-197 b, a hot Saturn orbiting an oscillating subgiant, in the first year of the \tess mission indicated promising prospects for a growing sample \citep{huber2019}. This was followed by the confirmation of TOI-257 b, a warm sub-Saturn orbiting an evolved F-type star \citep{addison2021}. \tess has also been very successful at recovering stellar oscillation signals for stars previously known to host planets. Asteroseismology of known planet hosts has been used to precisely characterize numerous planets orbiting evolved stars \citep{campante19,nielson2020,pepper2020,jiang2020,ball2020,hill2021,jiang2023}. \cite{huber2022} showed the power of the 20-second cadence observations from \tess to study pulsations in solar analogs, using the well studied $\pi$ Men system as a test case. Beyond these early discoveries and measurements in known planet hosts, there has been a surprising lack of newly discovered oscillating planet hosts from \tess.

The amplitude of p-mode oscillations increases as a star evolves \citep{chaplin13a}, making asteroseismic detections more likely for stars ascending the red giant branch. However, the increased radius and luminosity limit the likelihood of transit detection. The Giants Transiting Giants survey (GTG; \citealt{saunders2022,grunblatt2022,grunblatt2023,grunblatt2024,pereira2024,saunders2024b}) uses a pipeline developed to identify planets transiting the most evolved stars, and therefore provides an ideal sample to search for oscillating hosts. The precise constraints that asteroseismology enables can be used to test a variety of long-standing questions in exoplanet science, such as whether hot Jupiters are re-inflated at late times \citep{grunblatt16,grunblatt17,thorngren2021} and how the occurrence rate of giant planets changes as a function of stellar evolutionary state \citep{grunblatt19}.

Here, we present a new planet discovery and confirmation from the GTG survey---\planetname, a hot Saturn orbiting an oscillating low-luminosity red giant. We detect p-mode oscillations in the \tess light curve of \starname and perform asteroseismic modeling to constrain the stellar properties. 


\section{Observations} \label{sec:observations}

\subsection{\tess Photometry} \label{sec:tess}

We identified the transit signal of \planetname in the \tess Full-Frame Image (FFI) light curve produced by the \giants\footnote{\href{https://github.com/nksaunders/giants}{github.com/nksaunders/giants}} pipeline \citep{saunders2022}. The transit was initially flagged in a visual search and submitted as a Community \tess Object Of Interest (CTOI) in May, 2021. \starname has been observed in the \tess FFIs in Sectors 1 \& 2 at 30-minute cadence, 28 \& 29 at 10-minute cadence, and 68 \& 69 at 200-second cadence. This target additionally received 2-minute cadence observations during Sectors 68 \& 69. The full \tess observational baseline spans $\sim$1,883 days, from July 25, 2018 - September 20, 2023. The \textsf{giants} light curve used to identify \planetname was composed of data from the \tess FFIs for Sectors 1, 2, 28, and 29. We performed a Box-Least Squares (BLS) search for periodic signals using the \textsf{astropy.timeseries} implementation of the BLS method \citep{kovacs2002}. Full details about our search pipeline can be found in \cite{saunders2022}. We confirmed that the transit signal is detected in the FFI light curves produced by the TESS-SPOC \citep{caldwell2020} and QLP pipelines \citep{huang2020}, as well as the 2-minute cadence light curve produced by the SPOC pipeline \citep{jenkins2020}. Moreover, the difference image centroiding test \citep{twicken2018} performed on the 2-minute data constrained the location of the transit source to within $11.9\pm6.8$ arcsec of the host star.

\subsection{Radial Velocity Follow-up} \label{sec:rvs}

We performed ground-based radial velocity (RV) follow-up with three instruments to measure the mass and orbital eccentricity of \planetname. We obtained five observations of \starname with the CHIRON optical echelle spectrometer \citep{tokovinin2013} on the SMARTS 1.5m telescope at CTIO between June 30, 2023, and July 5, 2023. The median RV uncertainty of the CHIRON observations was 28.0 m s$^{-1}$.

We additionally obtained nine RV observations with the Carnegie Planet Finder Spectrograph (PFS; \citealt{crane2006,crane2008,crane2010}) on the 6.5m Magellan II telescope at Las Campanas Observatory in Chile. PFS is an optical echelle spectrograph with an iodine cell for wavelength calibration. Observations were obtained between May 26, 2024, and July 1, 2024. The median RV uncertainty of the PFS observations was 1.19 m s$^{-1}$.

Finally, we observed \starname with the fibre-fed FEROS spectrograph mounted on the MPG 2.2m \citep{Kaufer99} telescope at La Silla Observatory in Chile. These observations were performed in the context of the Warm gIaNts with tEss (WINE) collaboration \citep{wine0,wine1,wine2,wine3,wine4}. Twelve RVs were obtained between August 22, 2023, and August 19, 2024. FEROS spectra were obtained with the simultaneous calibration mode and were processed with the automatic \textsf{ceres} \citep{ceres} pipeline. The FEROS observations have the longest single-instrument baseline in our dataset, providing valuable information about the long-period RV variability. These data had a median RV uncertainty of 6.65 m s$^{-1}$. Table \ref{tab:rv_obs_table} in Appendix \ref{sec:rv_table} contains a full list of all radial velocity observations used in this work.



\section{Host Star Characterization} \label{sec:hoststars}

\subsection{High Resolution Spectroscopy} \label{sec:spectra}


To measure atmospheric parameters we used an out-of-transit, iodine-free template PFS spectrum. We restricted our analysis to the wavelength range between 500-630\,nm, for which the spectrum has a peak SNR of $\sim$\,200 at $\sim$ 580 nm. Continuum correction was performed by iteratively fitting 4th order polynomials to the 90th percentile flux for each spectral order binned into 20 wavelength segments. The resulting continuum-normalized spectrum was then analyzed using iSpec \citep{blanco14} to derive atmospheric parameters. We used the \textsf{turbospectrum} synthesis code \citep{plez12} with MARCS model atmospheres \citep{gustafsson08}, solar abundances from \citet{Grevesse07}, and the Gaia-ESO line list as implemented in iSpec \citep{blanco14}, excluding the sodium doublet. We fitted for T$_{\rm eff}$, $\log{g}$, [M/H] and \vsini, with microturbulence and macroturbulence parameters fixed using the built-in iSpec relations. The resulting best-fit yielded  T$_{\rm eff} = 4700$ K, $\log{g}=$ 3.2 dex, [M/H] = 0.25 dex, with no significant rotational broadening (<3 km s$^{-1}$). The temperature is in good agreement with photometric estimates from isochrone fitting (T$_{\rm eff}=4700$ K, \S \ref{sec:grid_modeling}) and the TESS Input Catalog (T$_{\rm eff}=4696$ K, \citealt{stassun2019}). The surface gravity is furthermore in good agreement with asteroseismology (see \S \ref{sec:grid_modeling}). We adopt uncertainties of 100 K in T$_{\rm eff}$ \citep[$\approx$\,2\%, following][]{tayar2020} and 0.1 dex in [M/H] \citep{torres12,furlan18} to account for possible systematic errors between different methods. 

\subsection{Asteroseismic Detection} \label{sec:seismology_modes}

We performed a search for stellar p-mode oscillations in the \tess light curve of \starname to provide additional constraints on stellar properties. First, we searched the \tess asteroseismic catalog produced by \cite{hon2021} and did not find reports of a seismic detection. Utilizing the \tess Asteroseismic Target List (\textsf{TESS-ATL}\footnote{\href{https://github.com/danhey/tess-atl}{github.com/danhey/tess-atl}}; \citealt{hey2024}) toolkit, we computed an asteroseismic detection probability for \starname of 100\%. We then performed an independent search for stellar oscillations. 

\begin{figure}[ht!]
    \centering
    \includegraphics[width=.45\textwidth]{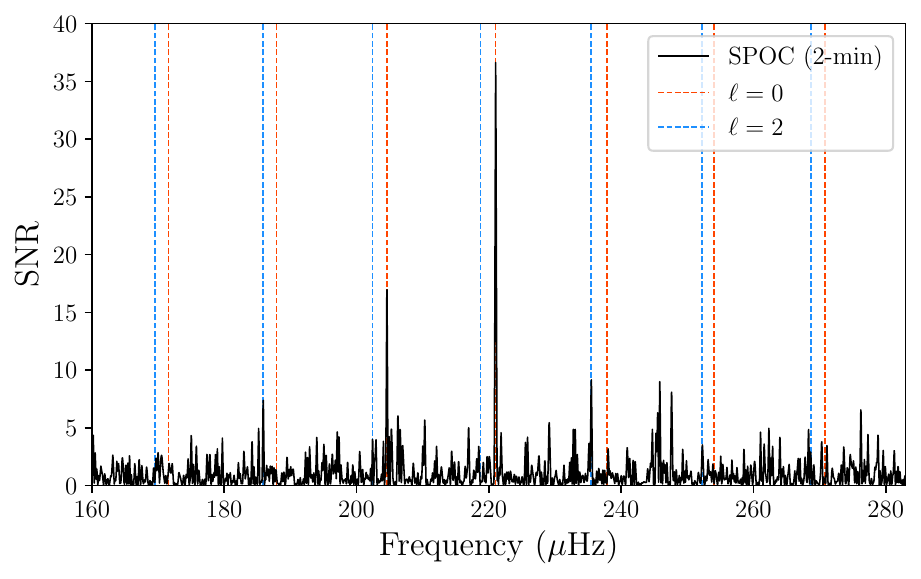}
    \caption{Amplitude spectrum of \starname centered on the range of frequencies showing stellar oscillations. The y-axis shows the signal-to-noise ratio of the oscillation power at each frequency, calculated by dividing the power spectrum by the estimated background. The orange and blue lines show the identified $\ell=0$ and $\ell=2$ modes, respectively.}
    \label{fig:tic201175570_numax}
\end{figure}

To produce our power spectrum we used the SPOC-generated Presearch Data Conditioning Simple Aperture Photometry (PDCSAP; \citealt{smith12,stumpe12,stumpe2014}) light curve composed of 2-minute cadence observations obtained in Sectors 68 and 69. PDCSAP light curves were selected as these data provided the highest signal-to-noise ratio (additional discussion in \S\ref{sec:future_seis}). We used the \lightkurve Python package \citep{lightkurve} to produce an amplitude spectrum and identified the frequency of maximum power, $\nu_{\rm max}$, using the 2D autocorrelation function method \citep{huber09,viani2019}. We identified a power excess with an envelope that peaks at a frequency of $\nu_{\rm max}\approx220.5$ $\mu$Hz with a large frequency separation of $\Delta\nu\approx16.45$ $\mu$Hz. Figure \ref{fig:tic201175570_numax} shows the envelope of oscillations identified in the \tess light curve, centered on the frequency of maximum power.

\begin{figure}[ht!]
    \centering
    \includegraphics[width=.48\textwidth]{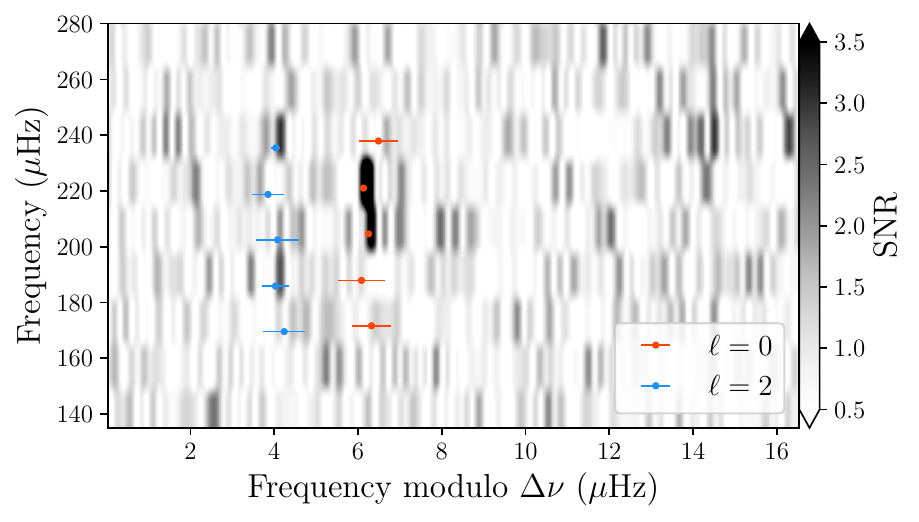}
    \caption{Frequency \'echelle diagram of the smoothed power spectrum. Orange points indicate the identified $\ell=0$ modes and blue points indicate the $\ell=2$ modes. The shading shows the signal-to-noise ratio at each frequency.}
    \label{fig:pbjam_echelle}
\end{figure}

We then performed seismic power spectrum modeling using the values of $\nu_{\rm max}$ and $\Delta\nu$ derived from the 2D autocorrelation as the initial values. We used the ``peakbagging" code \textsf{PBJam} \citep{nielsen_pbjam_2021} to identify pairs of $\ell=0,2$ modes in the power spectrum with the \textsf{PyMC3} \citep{exoplanet:pymc3} implementation of Hamiltonian Monte Carlo (HMC) sampling. \textsf{PBJam} then fit a Lorentzian profile to each mode and obtained the following constraints on the fundamental seismic parameters: $\nu_{\rm max}=$ \snumax $\mu$Hz, $\Delta\nu=$ \sdnu $\mu$Hz, and the $\ell=0,2$ separation, $\delta\nu_{02}=2.0906\pm0.0491$ $\mu$Hz. Figure \ref{fig:pbjam_echelle} shows the \'echelle diagram with modes identified by \textsf{PBJam}. The \'echelle diagram was produced by dividing the power spectrum into equal segments with length $\Delta\nu$ and stacking them vertically such that the $\ell=0$ and $\ell=2$ modes form ridges.

\subsection{Luminosity Constraint} \label{sec:isoclassify}

We used the \textsf{isoclassify}\footnote{\href{https://github.com/danxhuber/isoclassify}{https://github.com/danxhuber/isoclassify}} Python package \citep{huber2017,berger20} to compute the bolometric luminosity of \starname. We ran the code in direct mode, providing observables from our spectroscopic fit ($T_{\rm eff}$, [Fe/H]), the asteroseismic analysis ($\nu_{\rm max}$, $\Delta\nu$, $\log{g}$), Gaia DR3 (position and parallax distance; \citealt{gaiadr3}) as well as the K-band magnitude adopted from 2MASS \citep{skrutskie06}.
We used the Combined19 allsky dust map from \textsf{mwdust} \citep{bovy16}. From \textsf{isoclassify}, we report the luminosity, $L$, and distance, $d$ in Table \ref{tab:stellar}.

\subsection{Stellar Modeling} \label{sec:grid_modeling}

\begin{figure}[ht!]
    \centering
    \includegraphics[width=.45\textwidth]{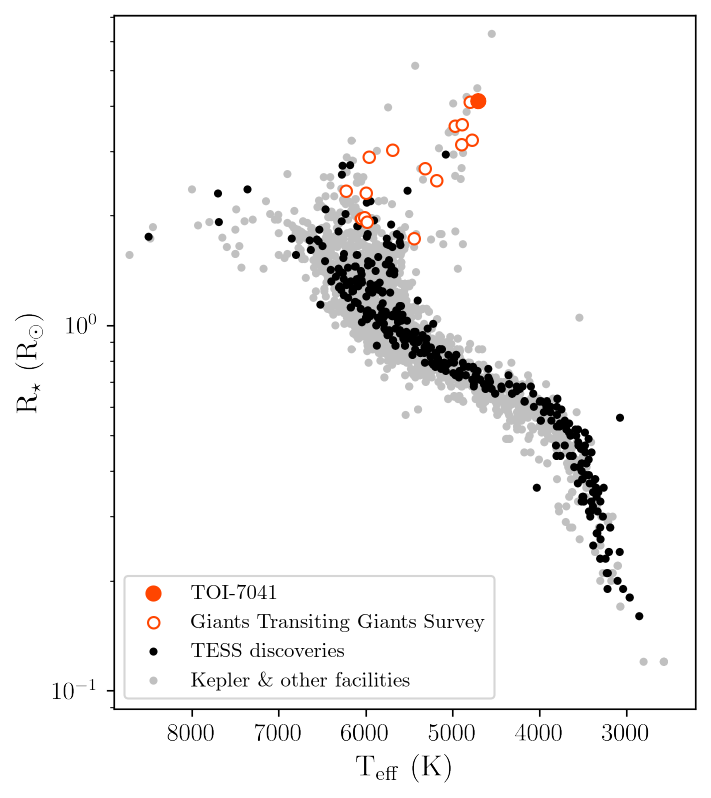}
    \caption{Stellar radius versus stellar effective temperature for all confirmed transiting planet hosts. \tess discoveries are shown in black and discoveries from other telescopes are shown in gray. Systems discovered by the GTG survey are marked by orange circles, with \starname indicated by the filled orange point.}
    \label{fig:hr}
\end{figure}

We calculated the stellar mass ($M_\star$), stellar radius ($R_\star$), surface gravity ($\log{g}$), and age using a model grid created with the Modules for Experiments in Stellar Evolution \citep[MESA\footnote{Details about the construction of the grid used in this work can be found in this GitHub repository: \href{https://github.com/parallelpro/mesa-rc-mass-loss}{github.com/parallelpro/mesa-rc-mass-loss}.};][]{paxton2010,paxton2013,paxton2015,paxton2018,paxton2019,jer23}. We used the measured values of $\nu_{\rm max}$, $\Delta\nu$, T$_{\rm eff}$, [M/H], and $\delta\nu_{02}$ to compute a likelihood function over the grid per the usual $\chi^2$ discrepancy statistic, which we convert to posterior probabilities under uninformative uniform priors on the stellar age by dividing out the sampling function of the grid. We further impose an additional prior constraint excluding stellar ages above $13.8\,\mathrm{Gyr}$ using a half-Gaussian cutoff function at that age, with $\sigma = 0.5\,\mathrm{Gyr}$. We report the posterior-weighted mean, and take the posterior-weighted standard deviation across the grid to be a measure of our statistical uncertainty. We repeat this exercise using the model grid of \cite{lindsay_subgiant_2024}, which was constructed using different model physics, and take the absolute difference between the posterior means reported by the two grids as an estimate of the systematic modelling uncertainty. The resulting asteroseismic quantities and stellar properties are reported in Table \ref{tab:stellar}. The stellar radius inferred from asteroseismology and the spectroscopic T$_{\rm eff}$ are shown in Figure \ref{fig:hr}.

\begin{table*}[]
    \centering
    \begin{tabular}{l l c c}
    \toprule
         & & \textbf{\starname} & Source \\
         \midrule
         TIC ID & & 201175570 & a  \\
         RA & & $23:51:12.52$ & a  \\
         dec & & $-50:52:11.5$ & a \\
         V Mag & & $11.251\pm0.026$ & a \\
         K Mag & & $8.691\pm0.023$ & a \\
         \textit{Gaia} Mag & & $10.9080\pm0.0002$ & a \\
         \textit{TESS} Mag & & $10.264\pm0.006$ & a \\
         T$_{\rm eff}$ & (K) & \teff & b \\ \relax 
         [M/H] & (dex) & \feh & b \\ 
         $v\sin{i_\star}$ & (km s$^{-1}$) & $<3$* & b \\ 
         M$_\star$ & (M$_\odot$) & \mstar & c \\
         R$_\star$ & (R$_\odot$) & \rstar & c \\ 
         $\log{g}$ & (dex) & \logg & c \\
         Age & (Gyr) & \starage & c \\
         L & (L$_\odot$) & $7.60^{+0.28}_{-0.26}$ & d \\
         $d$ & (pc) & $442\pm3$ & d \\
         $\nu_{\rm max}$ & ($\mu$Hz) & \snumax & e \\
         $\Delta\nu$ & ($\mu$Hz) & \sdnu & e \\
         $\delta\nu_{02}$ & ($\mu$Hz) & \sdzerotwo & e \\
         \bottomrule
    \end{tabular}
    \caption{Stellar properties derived for \starname. Sources: (a) \textit{TESS} input catalog v8.2 \citep{stassun2019}; (b) spectroscopic fit (this work); (c) asteroseismic grid-based modeling (this work); (d) \textsf{isoclassify} (this work; \citealt{huber2017,berger20}); (e) \textsf{PBJam} (this work; \citealt{nielsen_pbjam_2021}). *Upper limit only.}
    \label{tab:stellar}
\end{table*}

We performed an independent analysis to infer the fundamental and photospheric stellar parameters of \starname using the \textsf{isochrones} \citep{mor15} package to execute with \textsf{MultiNest} \citep{fer08,fer09,fer19} a simultaneous Bayesian fit of the MESA Isochrones \& Stellar Tracks \citep[MIST;][]{dot16,cho16} isochrone grid to a curated collection of data for the star.  We fit the MIST grid to the following photometric measurements: SkyMapper Southern Survey DR4 $uvgr$ photometry including in quadrature their zero-point uncertainties (0.03, 0.02, 0.01, 0.01) mag \citep{onk24}, Gaia DR2 $G$ photometry including in quadrature its zero-point uncertainty \citep{gai16,gai18,are18,bus18,eva18,rie18}, 2MASS $JHK_{s}$ photometry including their zero-point uncertainties \citep{skr06}, and WISE CatWISE2020 $W1W2$ photometry including in quadrature their zero-point uncertainties (0.032, 0.037) mag \citep{wri10,mai11,eis20,mar21}. We also fit to the $\Delta \nu$ and $\nu_{\text{max}}$ reported in Table \ref{tab:stellar}, a zero-point-corrected Gaia DR3 parallax \citep{gai21,fab21,lin21a,lin21b,row21,tor21}, and an estimated extinction value based on a three-dimensional extinction map \citep{lal22,ver22}.

As priors, we used a \citet{cha03} log-normal mass prior for $M_{\ast} < 1~M_{\odot}$ joined to a \citet{sal55} power-law prior for $M_{\ast} \geq 1~M_{\odot}$, a metallicity prior based on the Geneva-Copenhagen Survey \citep[GCS;][]{cas11}, a log-uniform age prior between 1 and 10 Gyr, a uniform extinction prior in the interval 0 mag $< A_{V} < 0.5$ mag, and a distance prior proportional to volume in the range of the \citet{bai21} geometric distance minus/plus five times its uncertainty.

The isochrone fit provides an estimate of effective temperature T$_{\rm eff}=4640\pm10$ K, surface gravity $\log{g}=3.24\pm0.01$, metallicity [Fe/H] = $0.39\pm0.02$, mass $M=1.12\pm0.02$ $M_\odot$, radius $R=4.20^{+0.03}_{-0.02}$ $R_\odot$, and age $t=9.4\pm0.4$ Gyr. These constraints are broadly consistent with the stellar properties inferred by spectroscopy and asteroseismology. The joint posterior distributions for our fit parameters can be found in Figure \ref{fig:iso_corner} in Appendix \ref{sec:iso_posteriors}. We adopt the asteroseismic measurements of mass, radius, $\log{g}$, and age, and the broader T$_{\rm eff}$ uncertainties from the spectroscopic analysis. 

\section{Planet Modeling} \label{sec:modeling}

\subsection{Simultaneous Transit \& RV Fitting} \label{sec:exoplanet}

We used the \exoplanet Python package \citep{exoplanet:exoplanet} to simultaneously fit an orbital model to the photometry and radial velocity observations. The data used in our model fit were the PDCSAP \tess photometry for Sectors 68 and 69 and the 26 RV observations listed in Table \ref{tab:rv_obs_table}. We parameterized eccentricity by optimizing the parameters $\sqrt{e}\sin{\omega}$ and $\sqrt{e}\cos{\omega}$ where $\omega$ is the argument of periastron. This parameterization avoids biasing the model towards higher eccentricities during sampling \citep{anderson2011, eastman2013}. In our model, eccentricity $e$ was bounded by $0 \leq e < 1$ and argument of periastron by $-\pi < \omega < \pi$. We use an eccentricity prior prescribed by the \cite{kipping2013a} Beta distribution. The other transit parameters we optimized were radius ratio $R_P/R_\star$, impact parameter $b$, orbital period $P$, and midtransit time at a reference epoch $t_0$. The radial velocity components were parameterized with a separate RV offset and jitter term for each of the three instruments. To estimate mass, we optimized the semi-amplitude $K$ of the RV trend. 

These distributions were created within a \textsf{PyMC3} model \citep{exoplanet:pymc3}, allowing us to optimize the model parameters using gradient descent. We sampled our model parameters using No U-Turn Sampling (NUTS; \citealt{nuts}) with four chains of 4,000 draws, with 4,000 iterations used to tune the model. 

\begin{figure*}[ht!]
\centering
\gridline{\fig{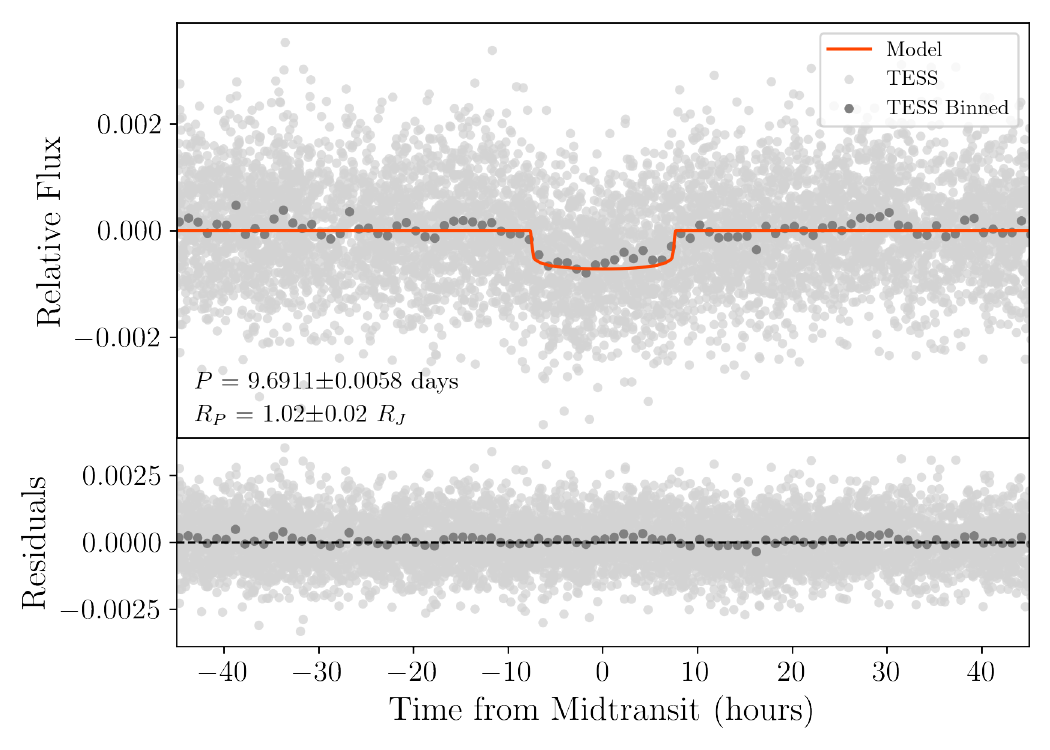}{.45\textwidth}{\textbf{(a)}}
          \fig{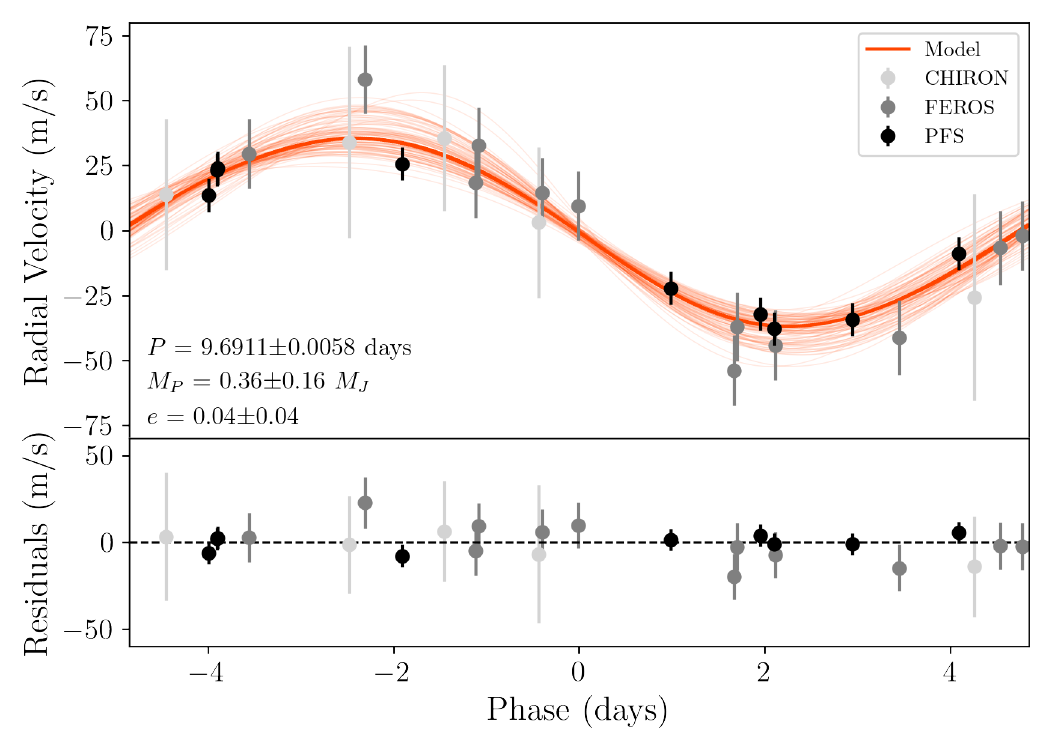}{.45\textwidth}{\textbf{(b)}}}
\caption{\textbf{(a)} Phase-folded \tess light curve, centered on the transit of \planetname. Light gray points show the \tess observations, dark gray points show \tess data binned to one hour, and the orange line shows the best-fit transit model. \textbf{(b)} RV measurements obtained with CHIRON (light gray), FEROS (dark gray), and PFS (black), phase-folded to the period of the transit signal. The orange line shows our best-fit orbital model, with random draws from the posterior distribution of the model shown by the fainter orange lines. The RV and transit models were fit simultaneously using the \exoplanet Python package.}
\label{fig:201175570_fits}
\end{figure*}

We report our fit results in Table \ref{tab:fit_results}, adopting the median value of the posterior distribution for each parameter and its standard deviation as the uncertainty. Our resulting orbital models can be found in Figure \ref{fig:201175570_fits}, which shows the transit model fit (left) and the radial velocity solution (right).

\begin{table*}[]
    \centering
    \begin{tabular}{l l c}
    \toprule
         & & \textbf{TOI-7041 b} \\
         \midrule
         \multicolumn{2}{l}{\textit{Fitted parameters:}} \\
$P$ & (days) & \porb  \\
$t_0$ & (BJD) & $2460160.393\pm0.012$  \\
$R_p/R_\star$ & & \rprs  \\
$a/R_\star$ & & $4.8\pm0.1$  \\
$b$ & & \impactparam  \\
$K$ & (m/s) & \rvk \\
$\omega$ & ($^\circ$) & \plomega \\
$e$ & & \ecc \\
\multicolumn{2}{l}{\textit{Derived parameters:}} \\
$M_p$ & ($M_J$) & \mplanet  \\
$R_p$ & ($R_J$) & \rplanet  \\
         \bottomrule
    \end{tabular}
    \caption{Best-fit orbital parameters for \planetname.}
    \label{tab:fit_results}
\end{table*}

\subsection{Search for Additional Planets} \label{sec:additional_planets}

We searched for additional transiting planets in the \tess photometry by masking out the transits of \planetname and performing a BLS search on the resulting light curve. We searched a grid of 10,000 periods between 1 and 50 days and 1,000 durations between 2 hours and 20 hours. No periodic signals were identified above a signal-to-noise ratio of 10.

We also searched for signatures of non-transiting planets in the RVs. Using the \textsf{RVSearch}\footnote{\href{https://github.com/California-Planet-Search/rvsearch}{github.com/California-Planet-Search/rvsearch}} Python package, we calculated a Lomb-Scargle periodogram and searched the result for evidence of a single most-significant planet, then iteratively searched for additional periodic components in the RV time series. We identify an additional periodic signal at a period of $149.5\pm0.1$ days with an RV semi-amplitude of $K=55\pm6$ m s$^{-1}$. We searched the \tess light curve for transits at times that would be consistent with this periodic RV component, and did not identify a transit in the single such time that occurred during a \tess observation.

\begin{figure}[ht!]
    \centering
    \includegraphics[width=.45\textwidth]{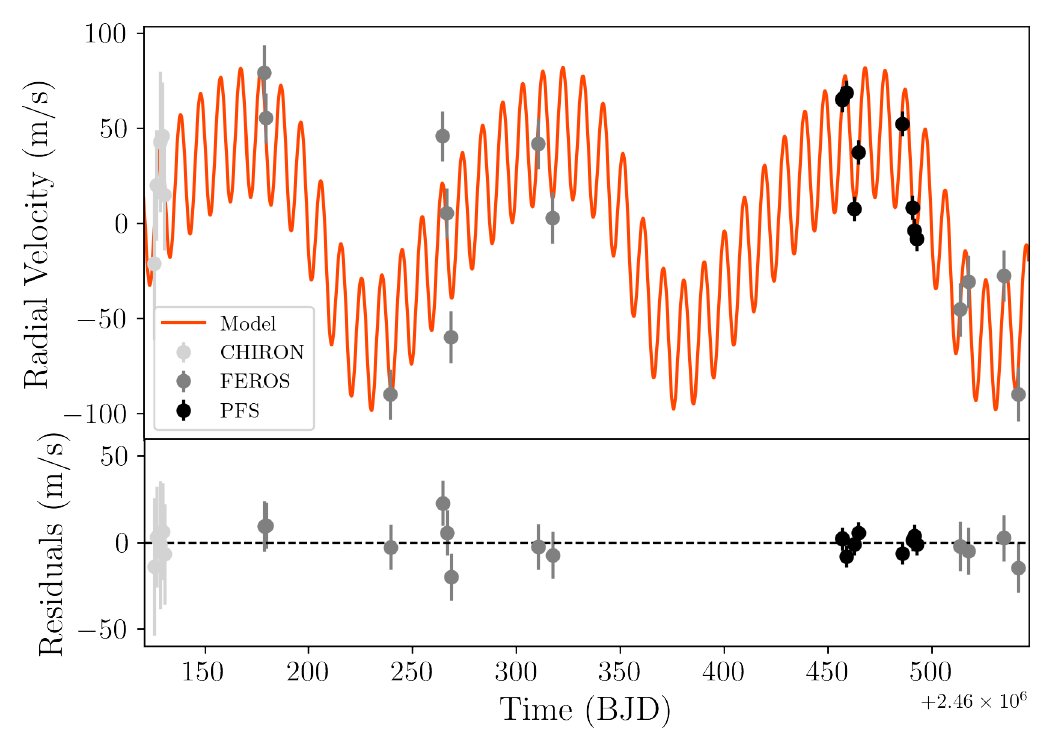}
    \caption{Full RV timeseries for \starname with our two-planet model shown in orange. In addition to the signal in-phase with the transit ephemeris (shown as a phase-folded RV curve in Figure \ref{fig:201175570_fits}, right), we identify a periodic trend with a period of $\sim$150 days.}
    \label{fig:rv_full}
\end{figure}

Assuming the signal is planetary in origin and the orbit is observed near edge-on, this amplitude would correspond to a planet mass of $\sim$0.6 $M_J$. We perform our fitting routine with the inclusion of this trend, and the best-fit model reports a moderate eccentricity for the potential outer planet of $e=0.23\pm0.07$. Our two-planet model is shown with the full RV timeseries in Figure \ref{fig:rv_full}. The additional signal has been removed from the phase-folded RV model for the transiting planet, \planetname, shown in Figure \ref{fig:201175570_fits}, right.

Due to the limited baseline of our observations relative to the measured period of the additional RV signal, further monitoring of this system is required to confirm that the variability is caused by a planetary companion. Stellar activity signals have been shown to produce false positive detections in radial velocity observations, due to star spot modulation and magnetic or chromospheric variability \citep{saar1997,hatzes2018,delgatomena2018,simpson2022}. These false positive cases can be identified through correlations between RV variability and flux variability in the photometric time series. However, we do not identify this long-period variability in the \tess light curve. Continued RV monitoring will reveal whether this signal is coherent over long timescales, which could rule out the false positive case.

\section{Results} \label{sec:results}

\planetname is a hot Saturn ($R_p=$ \rplanet $R_J$, $M_p=$ \mplanet $M_J$) on a \porb day orbit around an oscillating red giant star. Our orbital model indicates that the planet's orbit is nearly circular ($e=$ \ecc). By analyzing the asteroseismic signal observed in the \tess light curve of \starname, we obtain estimates of the stellar mass $M_\star=$ \mstar $M_\odot$, stellar radius $R_\star=$ \rstar $R_\odot$, surface gravity $\log{g}=$ \logg dex, and age $t=$ \starage Gyr. \starname has an effective temperature of $T_{\rm eff}=$ \teff K, which when considered along with its mass and radius indicates that the star is a red giant. 


\section{Discussion} \label{sec:discussion}

\subsection{Comparison to Known Exoplanets} \label{sec:pop_compare}

\begin{figure}[ht!]
    \centering
    \includegraphics[width=.45\textwidth]{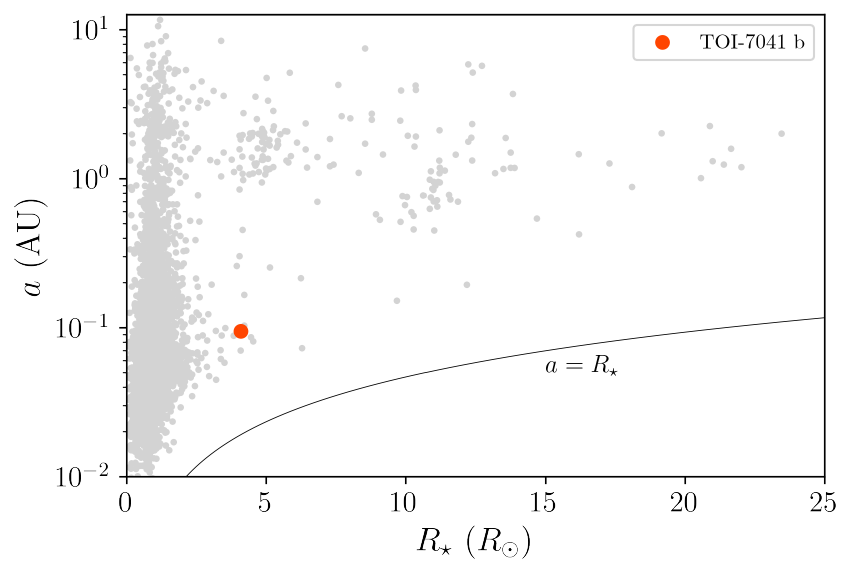}
    \caption{Planetary orbital semi-major axis ($a$) shown as a function of stellar radius ($R_\star$) for all confirmed exoplanet (gray) and \planetname (orange). The solid black line shows the 1:1 relation where $a=R_\star$.}
    \label{fig:a_Rstar}
\end{figure}

\starname may be the largest star with a confirmed planet discovered in \tess data. It is similar in both size and temperature (within 1-$\sigma$ of each) to TOI-2669 (GTG II; \citealt{grunblatt2022}). The position of \starname on a Hertzsprung-Russell diagram can be found in Figure \ref{fig:hr}, where it may be compared to the population of \tess-discovered host stars. Hosts of confirmed planets from the GTG survey make up the majority of subgiant and red giant hosts from \tess. The confirmation of \planetname moves us closer to the largest planet hosts from \kepler---Kepler-91 \citep{lillobox14} and Kepler-56 \citep{steffen12}---and \ktwo---K2-97 and K2-132 \citep{grunblatt16,grunblatt17,jones2018}. Figure \ref{fig:a_Rstar} shows the orbital semi-major axis, $a$, as a function of stellar radius, $R_\star$. Due to the large radius of its host, \planetname is positioned in the lower envelope of the distribution, near the line indicating the stellar surface ($a=R_\star$). This system is very near Kepler-56, K2-132, K2-97, and TOI-2669 in $a$ versus $R_\star$ space, while the closest point to the 1:1 line is Kepler-91.

\begin{figure}[ht!]
    \centering
    \includegraphics[width=.45\textwidth]{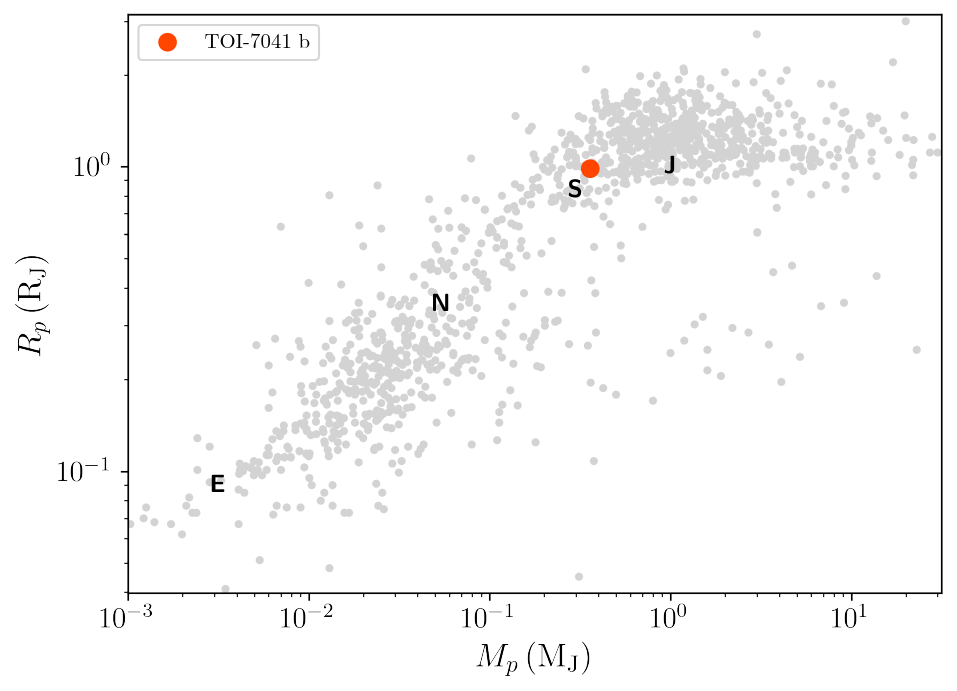}
    \caption{Planet radius versus planet mass for all known exoplanets are shown in gray. The position of \planetname is marked by the orange point. Solar system planets Earth, Neptune, Saturn, and Jupiter are marked by their initials.}
    \label{fig:mass_rad}
\end{figure}

Figure \ref{fig:mass_rad} shows the position of \planetname relative to all other confirmed exoplanets in radius versus mass. The planet's mass is close to that of Saturn (marked by the `S') and its radius is near 1 $R_J$.

\subsection{Planet Radius Re-inflation} \label{sec:reinflation}

The anomalously large radii of highly-irradiated giant planets is a long-standing mystery in exoplanet science \citep{demory2011,laughlin11,miller11,hartman2016}, with important implications for our understanding of planet interior physics. Theoretical predictions have indicated that giant planets on short orbital periods should undergo rapid re-inflation as their host stars brighten on the main sequence or as they become evolved \citep{thorngren2021}. However, the \tess sample of hot Jupiters orbiting evolved stars does not seem to follow as clear a radius-mass-flux relationship as the main sequence population, instead displaying a wider range of radii at high incident fluxes \citep{grunblatt2022}. Discoveries from \tess have also shown that lower-mass planets may be able to retain their atmospheres at higher incident fluxes than previously expected \citep{grunblatt2024}.

We examined how the incident flux received by \planetname changed as its host star evolved. To estimate the incident flux received by the planet on the main sequence, we produced a stellar model with MESA. Using the stellar properties listed in Table \ref{tab:stellar}, we initiated a stellar model and ran it from the pre-main sequence to the base of the red giant branch. We then identified the main sequence effective temperature and radius and computed the incident flux. On the main sequence, \planetname likely received incident flux below the threshold for inflation of $\sim$150 $F_\oplus$ defined in \cite{demory2011}. At this level of incident flux, a radius of $\sim$1 $R_J$ would not be inconsistent with the main sequence population, though it places \planetname among the largest planets of similar mass. The measured radius of \planetname indicates that it may have undergone moderate re-inflation as its host star evolved, but it is not significantly larger than systems which have not undergone a similar increase in incident flux.

\begin{figure}[ht!]
    \centering
    \includegraphics[width=.45\textwidth]{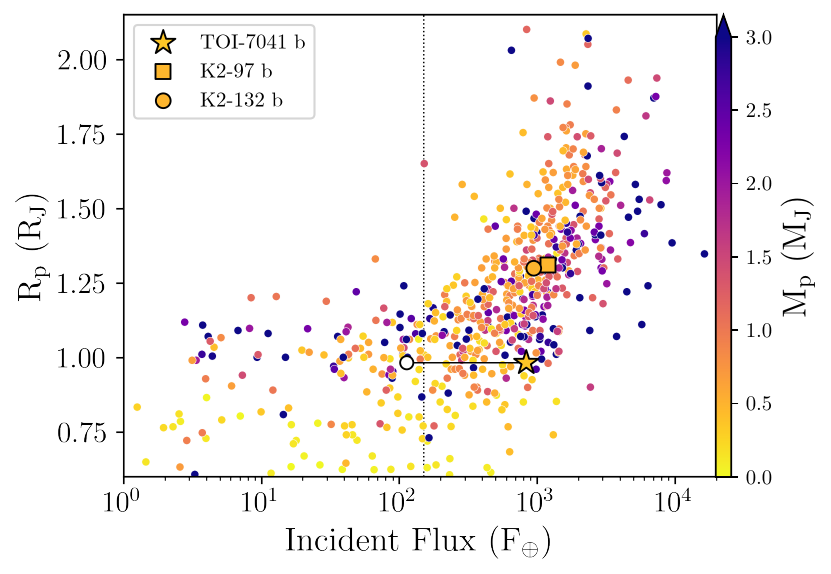}
    \caption{Planet radius shown as a function of incident flux the planet receives from its star. Point color indicates the planet mass. The dashed vertical line shows the inflation threshold from \cite{demory2011}. The estimated main sequence incident flux received by \planetname is indicated by the white circle.}
    \label{fig:flux_rad}
\end{figure}

Here, we compare \starname to two analog systems: K2-97 and K2-132. These systems are remarkably similar to \starname, being composed of a low-luminosity red giant host star ($T_{\rm eff}\approx4800$, $R_\star\approx4$ $R_\odot$) with a roughly Saturn-mass ($\sim$0.5 $M_J$) planet orbiting with a period of $\sim$9 days. We plot planet radius as a function of incident flux received from the host star in Figure \ref{fig:flux_rad}. \planetname sits on the lower edge of the inflation trend, though its radius is consistent with other planets of similar mass, which span a wide range of radii in a similar range of incident flux ($\sim10^3$ $F_\oplus$). When compared to the other evolved systems with similar planetary properties we have highlighted, \planetname is significantly smaller. Both K2-97 and K2-132 show substantial inflation, with radii near $\sim$1.3 $R_J$, compared to $\sim$1 $R_J$ for \planetname. 

The incident flux received by \planetname is similar to that of K2-97 b and K2-132 b, and we must therefore consider additional heating mechanisms to explain the difference in observed inflation. Tidal circularization of a planet's orbit can result in heat deposited deep in the planet's interior through tidal distortion \citep{bodenheimer01}. We measure a low eccentricity for \planetname ($e=$ \ecc), indicating that the planet is not undergoing tidal heating due to circularization. Conversely, K2-97 b and K2-132 b show significant non-zero eccentricities of $e=$ 0.22$\pm$0.08 and 0.36$\pm$0.06, respectively \citep{grunblatt18}. Given that eccentricity is the most significant distinguishing factor between these two inflated systems and the less inflated planet reported in this work, it appears that heating due to tidal circularization may be a dominant source of late-stage planetary radius re-inflation. 

\subsection{Prospects for Future Asteroseismic Detections} \label{sec:future_seis}

\starname was observed in six \tess sectors, which included 30-minute, 10-minute, and 2-minute cadence observations. The oscillation signature is clearly visible in the 2-minute cadence observations, weakly visible in the 10-minute cadence observations, and not detected in the 30-minute cadence observations. The peak frequency of the envelope of oscillations (\snumax $\mu$Hz) is below the Nyquist limit for each of these cadences and should therefore be detectable, though the amplitudes are likely undergoing Nyquist attenuation at the longest cadence, which has a Nyquist limit of 283 $\mu$Hz. The primary contributor to the difference in the recovered signal-to-noise ratio of the oscillations between 10- and 2-minute cadence sectors is likely the light curve de-trending.

For the asteroseismic analysis reported in this work, we use only the 2-minute cadence PDCSAP light curve produced by the SPOC pipeline \citep{jenkins2016}. We also performed the analysis using the combined amplitude spectra of the 30-minute, 10-minute, and 2-minute cadence light curves; however, we found that the inclusion of longer cadences resulted in an increasingly reduced SNR. Our longer-cadence light curves were produced using a variety of publicly available FFI pipelines, including TESS-SPOC \citep{caldwell2020}, QLP \citep{huang2020}, \textsf{eleanor-lite} \citep{feinstein2019}, TGLC \citep{han2023}, and our own \textsf{giants} pipeline. The recovery of oscillations with the highest signal-to-noise ratio in the 2-minute cadence SPOC light curve indicates that the de-trending applied by this pipeline preserves the oscillations and results in the lowest signal dilution from systematics or contaminating sources. \cite{huber2022} showed that the details of how observations are processed and reduced are crucial to extracting the highest-quality oscillation signals at 20-second cadence, and similar effects may be true for other cadences. The signal should still be measurable in the longer-cadence FFI observations, and the retrieval of such asteroseismic signals may become possible as existing pipelines are adapted in the future. However, the results for \starname indicate that obtaining targeted short-cadence light curves which are processed by the SPOC pipeline may enable improved asteroseismic analysis.

A larger sample of asteroseismic measurements for evolved planet hosts would provide valuable information for investigations of key open questions related to the evolution of planetary systems. Improved radius precision will be particularly valuable to test planet radius inflation scenarios. Discoveries from the GTG survey have already provided valuable benchmarks, and precise measurements of the planets' properties from asteroseismology of their hosts are crucial to better test theoretical predictions for planetary atmosphere inflation and interior physics. The occurrence of giant planets as a function of stellar evolutionary state is also unclear. Some studies have suggested that giant planets should be depleted by tides before the star becomes a red giant \citep{hamer2019}, while others have indicated that these planets survive to at least the base of the red giant branch and have a similar rate of occurrence to their main sequence counterparts \citep{grunblatt2019}. A large sample of evolved planet hosts with asteroseismic mass measurements would enable a direct comparison between similar stars at varied evolutionary states, providing a clearer look at the time dependence of planet occurrence.

\section{Conclusions} \label{sec:conclusions}

Our main conclusions are:
\begin{enumerate}
    \item \planetname is a hot Saturn with a radius of $R_p=$ \rplanet $R_J$ on a roughly circular orbit ($e=$ \ecc) with an orbital period of $P=$ \porb days.
    \item Radial velocity observations show tentative evidence of an outer companion to \planetname with an orbital period of $\sim$150 days. Further monitoring of this system is required to distinguish this trend from a stellar signal and constrain the orbit of the potential companion.
    \item \planetname is significantly less inflated than similar systems (K2-97 b and K2-132 b), despite the similarities in the planet masses and incident fluxes. The low eccentricity measured for \planetname, when compared to the more eccentric orbits of K2-97 b and K2-132 b, indicates that the system is not undergoing tidal circularization, and points to tidal heating as a potential source of planet radius re-inflation.
    \item We measured solar-like oscillations in the \tess light curve of \starname that peak near $\nu_{\rm max}=$ \snumax $\mu$Hz and have a large frequency separation of $\Delta\nu=$ \sdnu $\mu$Hz. This system joins a small number of \tess-discovered planets with asteroseismically-characterized host stars, and is the first oscillating red giant host from \tess.
    \item We performed grid-based asteroseismic modeling of the observed oscillation signal to infer precise stellar properties. \starname is one of the largest \tess stars known to host a transiting exoplanet, with a radius of $R_\star=$ \rstar $R_\odot$. We report an age for the system of $t=$ \starage Gyr.
    \item Stellar oscillations are observed with the highest signal-to-noise ratio in the 2-minute cadence observations processed by the SPOC pipeline, indicating that we will likely recover additional asteroseismic detections as more targeted short-cadence observations of evolved stars are performed.
\end{enumerate}

Despite a small number of asteroseismic planet hosts from \tess to date, the GTG survey has produced a large sample of promising targets for future asteroseismic analysis. The discovery and precise characterization of \planetname and similar systems will allow us to study in detail the changes that planetary systems undergo as their host stars evolve.

\section*{Acknowledgements}\label{sec:acknowledgements}

N.S. acknowledges support by the National Science Foundation Graduate Research Fellowship Program under Grant Numbers 1842402 \& 2236415 and the National Aeronautics and Space Administration (80NSSC21K0652). 
D.H. acknowledges support from the Alfred P. Sloan Foundation, the National Aeronautics and Space Administration (80NSSC21K0652), and the Australian Research Council (FT200100871).
J.M.J.O. acknowledges support from NASA through the NASA Hubble Fellowship grant HST-HF2-51517.001, awarded by STScI. STScI is operated by the Association of Universities for Research in Astronomy, Incorporated, under NASA contract NAS5-26555.
R.B. acknowledges support from FONDECYT Project 1241963 and from ANID -- Millennium  Science  Initiative -- ICN12\_009.
A.J. acknowledges support from ANID -- Millennium Science Initiative -- ICN12\_009, AIM23-0001 and from FONDECYT project 1210718.
D.R. was supported by NASA under award number NNA16BD14C for NASA Academic Mission Services.
T.T. acknowledges support by the BNSF program ``VIHREN-2021'' project No. KP-06-DV/5.

This paper made use of data collected by the TESS mission and are publicly available from the Mikulski Archive for Space Telescopes (MAST) operated by the Space Telescope Science Institute (STScI). Funding for the TESS mission is provided by NASA's Science Mission Directorate. We acknowledge the use of public TESS data from pipelines at the TESS Science Office and at the TESS Science Processing Operations Center. Resources supporting this work were provided by the NASA High-End Computing (HEC) Program through the NASA Advanced Supercomputing (NAS) Division at Ames Research Center for the production of the SPOC data products.

\software{\textsf{Lightkurve} \citep{lightkurve}, \textsf{Astropy} \citep{astropy2013,astropy2018,astropy2022}, \textsf{Astroquery} \citep{astroquery}, \textsf{PBJam} \citep{nielsen_pbjam_2021}, \textsf{echelle} \citep{echelle}, \textsf{isochrones} \citep{mor15}, \textsf{R} \citep{r24}.}

\pagebreak
\nocite{tange2018}
\bibliography{references,references2}{}
\bibliographystyle{aasjournal}

\appendix

\section{Isochrone Fitting Posteriors} \label{sec:iso_posteriors}

\begin{figure*}[ht!]
    \centering
    \includegraphics[width=0.75\linewidth]{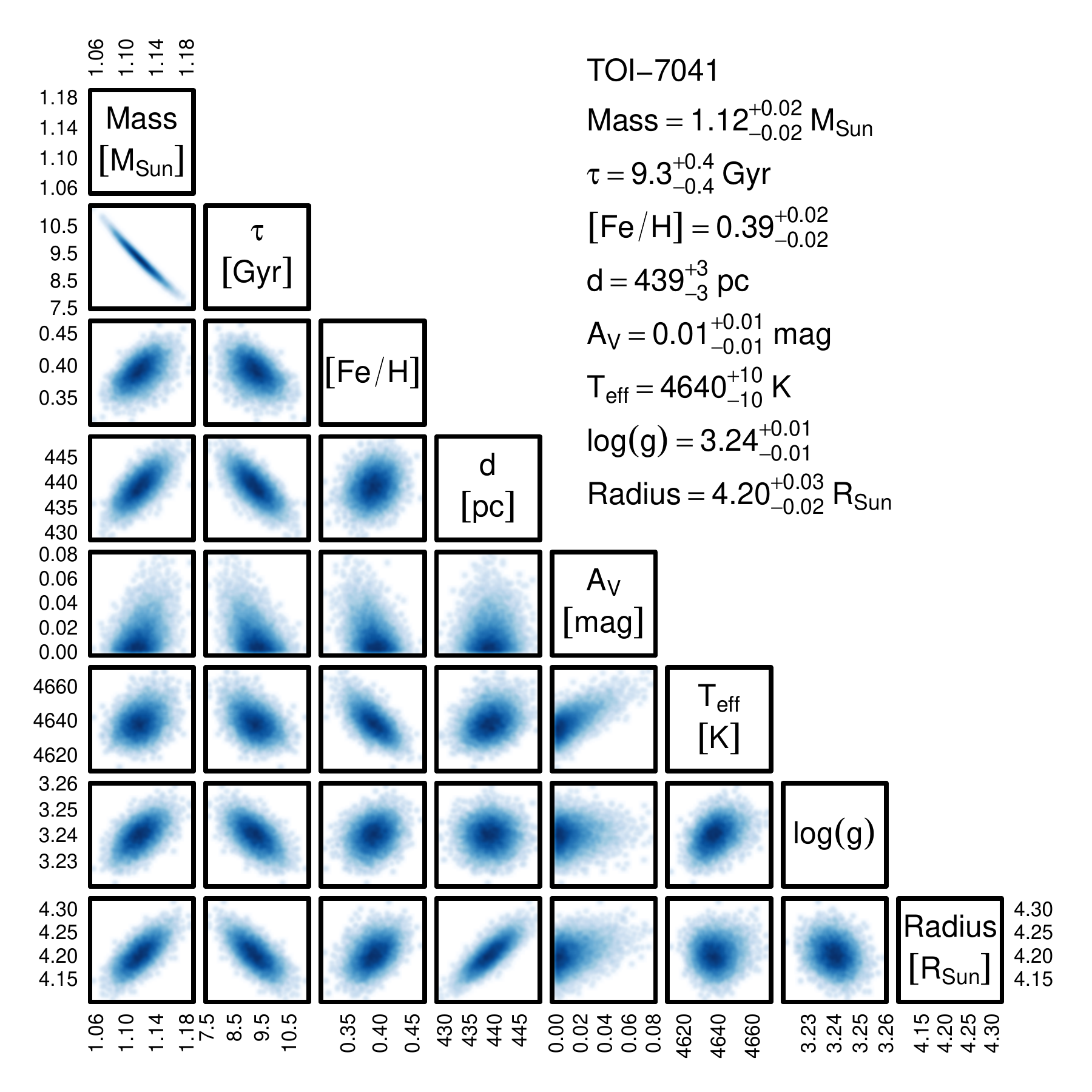}
    \caption{Joint posterior distributions from our isochrone grid modeling of \starname.}
    \label{fig:iso_corner}
\end{figure*}

In Figure \ref{fig:iso_corner}, we show the joint posterior distributions for the stellar properties fit with our isochrone grid. A detailed description of the analysis can be found in \S \ref{sec:grid_modeling}.

\section{Radial Velocity Observations} \label{sec:rv_table}

Table \ref{tab:rv_obs_table} lists all RV observations used in this analysis.

\begin{table}[ht!]
    \centering
    \begin{tabular}{l r r r}
    \toprule
    \textbf{Instrument} & \textbf{Time (BJD)} & \textbf{RV (m/s)} & \textbf{RV Error (m/s)} \\
    \midrule
CHIRON & 2460125.886770 & -21 & 39 \\
CHIRON & 2460126.869260 & 20 & 28 \\
CHIRON & 2460128.846650 & 43 & 36 \\
CHIRON & 2460129.867900 & 46 & 27 \\
CHIRON & 2460130.884460 & 15 & 28 \\
FEROS & 2460178.691553 & 79.1 & 8.9 \\
FEROS & 2460179.764638 & 55.3 & 6.2 \\
FEROS & 2460268.663061 & -59.8 & 6.9 \\
FEROS & 2460264.689207 & 45.8 & 6.1 \\
FEROS & 2460266.595236 & 5.3 & 6.1 \\
FEROS & 2460239.621033 & -89.9 & 6.1 \\
FEROS & 2460310.538160 & 41.8 & 6.3 \\
FEROS & 2460317.564912 & 2.9 & 6.8 \\
FEROS & 2460513.811147 & -45.3 & 8.0 \\
FEROS & 2460517.848128 & -30.7 & 7.2 \\
FEROS & 2460534.788253 & -27.6 & 6.5 \\
FEROS & 2460541.796333 & -89.9 & 8.2 \\
PFS & 2460456.913340 & 65.02 & 1.62 \\
PFS & 2460456.922440 & 65.59 & 1.59 \\
PFS & 2460458.907970 & 68.67 & 1.16 \\
PFS & 2460462.916010 & 7.55 & 1.19 \\
PFS & 2460464.909460 & 37.50 & 1.09 \\
PFS & 2460485.896450 & 52.39 & 1.56 \\
PFS & 2460490.879490 & 8.16 & 1.13 \\
PFS & 2460491.843090 & -3.79 & 1.19 \\
PFS & 2460492.832010 & -8.07 & 1.15 \\
\bottomrule
    \end{tabular}
    \caption{Radial velocity observations used in this analysis. The instrumental offset for each observation has been subtracted from the reported RV value.}
    \label{tab:rv_obs_table}
\end{table}

\end{CJK*}
\end{document}